\UseRawInputEncoding
\documentclass[journal,onecolumn]{IEEEtran}
\usepackage{amsmath,amsfonts}
\usepackage[ruled, linesnumbered]{algorithm2e}
\usepackage{array}
\usepackage{caption}
\usepackage{xcolor}
\usepackage{graphicx}
\usepackage[hidelinks]{hyperref}
\usepackage{subcaption}
\usepackage{multirow}
\usepackage{url}
\usepackage{graphicx}
\usepackage{lipsum}

\begin{document}

\title{HMPC-assisted Adversarial Inverse Reinforcement Learning for Smart Home Energy Management}


\author{
  {Jiadong~He},
  {Liang~Yu,~\IEEEmembership{Senior~Member,~IEEE}},
  {Zhiqiang~Chen},
  {Dawei~Qiu,~\IEEEmembership{Member,~IEEE}},
  {Dong~Yue,~\IEEEmembership{Fellow,~IEEE}},
  {Goran~Strbac,~\IEEEmembership{Senior~Member,~IEEE}},
  {Meng~Zhang,~\IEEEmembership{Senior~Member,~IEEE}},
  {Yujian~Ye,~\IEEEmembership{Senior~Member,~IEEE}},
  {Yi~Wang,~\IEEEmembership{Senior~Member,~IEEE}}
  \thanks{
    \newline J. He, L. Yu, Z. Chen, and D. Yue are with the College of Automation $\&$ College of Artificial Intelligence, Nanjing University of Posts and Telecommunications, Nanjing 210023, China. (e-mail: jiadonghe007@outlook.com, liang.yu@njupt.edu.cn, zqchen19981115@163.com, and medongy@vip.163.com)
    \newline D. Qiu is with the Department of Engineering, University of Exeter, Exeter, EX4 4PY, U.K. (email: d.qiu@exeter.ac.uk)
    \newline G. Strbac is with the Department of Electrical and Electronic Engineering, Imperial College London, London SW7 2AZ, U.K. (e-mail: g.strbac@imperial.ac.uk)
    \newline M. Zhang is with the School of Cyber Science and Engineering, Xi'an Jiaotong University, Xi'an 710049, China (e-mail:
    mengzhang2009@xjtu.edu.cn)
    \newline Y. Ye is with the School of Electrical Engineering, Southeast University, Nanjing, Jiangsu 210096, China. (e-mail: yeyujian@seu.edu.cn)
    \newline Y. Wang is with the Department of Electrical and Electronic Engineering, The University of Hong Kong, Hong Kong SAR, China (e-mail: yiwang@eee.hku.hk).
  }
}

\maketitle

\begin{abstract}
This letter proposes an Adversarial Inverse Reinforcement Learning (AIRL)-based energy management method for a smart home, which incorporates an implicit thermal dynamics model. In the proposed method, historical optimal decisions are first generated using a neural network-assisted Hierarchical Model Predictive Control (HMPC) framework. These decisions are then used as expert demonstrations in the AIRL module, which aims to train a discriminator to distinguish expert demonstrations from transitions generated by a reinforcement learning agent policy, while simultaneously updating the agent policy that can produce transitions to confuse the discriminator. The proposed HMPC-AIRL method eliminates the need for explicit thermal dynamics models, prior or predictive knowledge of uncertain parameters, or manually designed reward functions. Simulation results based on real-world traces demonstrate the effectiveness and data efficiency of the proposed method.
\end{abstract}


\section{Introduction}\label{s1}

\IEEEPARstart{B}{uilding} sector consumed approximately 30\% of global final energy and contributed to 26\% of energy-related carbon emissions in 2022. Among these, residential buildings represent a substantial share, accounting for nearly 60\% of total building energy consumption~\cite{GABC2022}. One of the most effective strategies to reduce energy use in residential buildings is the development of smart homes~\cite{Pinthurat2024}. In a smart home, the home energy management system (HEMS) serves as a core component, collecting operational data from devices (including photovoltaic (PV) panels, energy storage systems (ESSs), heating, ventilation, and air conditioning (HVAC) systems, and various household appliances) and making intelligent decisions, leading to lower energy consumption/cost, occupant discomfort, and renewable energy waste. Thus, it is of great importance to develop advanced smart home energy management (SHEM) methods.

Many SHEM approaches have been developed in existing works, which can be classified into model-based and data-driven approaches. Model-based approaches rely on explicit building thermal dynamics models and require prior or predictive information about uncertain parameters. However, due to complex factors (e.g., building structure, materials, and surrounding environment), it is challenging to obtain an accurate and explicit thermal dynamics model. Moreover, parameter prediction inevitably introduces errors, which can affect the performance of such approaches. To eliminate the above-mentioned reliance and requirements, deep reinforcement learning (DRL)-based SHEM approaches have been adopted~\cite{Pinthurat2024,yu2023}. Although such DRL-based approaches are effective, they require manually designed reward functions, which introduces complexity in selecting appropriate reward structures and parameters. To avoid such manual burden, Dey et al. employed maximum entropy inverse reinforcement learning (MaxEnt IRL) with rule-based demonstrations for building comfort control~\cite{dey2023inverse}. However, the MaxEnt IRL-based control method faces the issue of reward ambiguity (i.e., multiple reward functions may induce the same optimal policy), which can affect the robustness of policy learning~\cite{fu2017learning}. Moreover, rule-based expert demonstrations typically suffer from low quality.

To address the above limitations, this letter proposes a novel SHEM method based on historical optimization decision-driven Adversarial Inverse Reinforcement Learning (AIRL) under an implicit thermal dynamics model. Specifically, a neural network-assisted hierarchical model predictive control (HMPC) framework is employed to generate high-quality historical optimization decisions. These decisions are then used as expert demonstrations for the AIRL module, which is designed to train a discriminator to distinguish expert demonstrations from transitions of the HEMS agent policy, while simultaneously updating the agent’s policy to generate transitions that can confuse the discriminator. Simulation results demonstrate the effectiveness and data efficiency of the proposed method. To the best of our knowledge, this is the first work that applies HMPC-AIRL to SHEM.

\section{Problem Formulation}\label{s2}

\begin{figure}[ht]
  \centering
  \includegraphics[width=0.65\columnwidth]{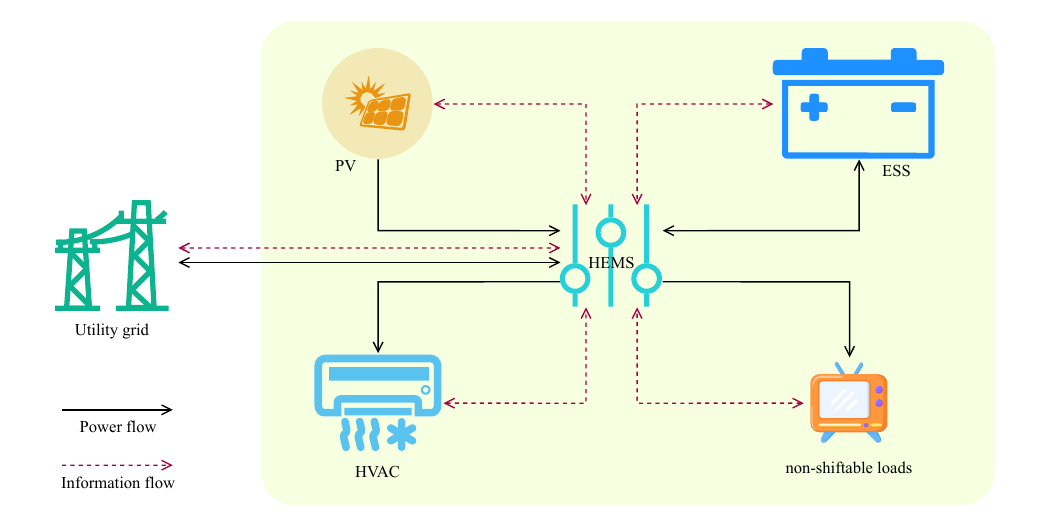} 
  \caption{The considered smart home energy system in this letter}
  \label{fig:system}
\end{figure}

This letter considers a smart home energy system as shown in Fig.~\ref{fig:system}, where PV, ESS, HVAC, non-shiftable loads, and an HEMS could be identified. The purpose of the HEMS is to minimize the long-term system energy cost while maintaining thermal comfort for occupants by controlling HVAC and ESS jointly. Suppose that the HEMS operates in slotted time, i.e., $t \in [1,M]$, where $M$ denotes the total number of time slots. Then, a long-term multi-objective optimization problem related to the HMES can be formulated by
\begin{gather}
\min_{e_t, h_t}\mathop {\lim \sup }\limits_{M \to \infty } \frac{1}{M}\mathbb{E}\Big\{\sum\nolimits_{t=1}^M(X_{1,t}+X_{2,t})\Big\} \label{eq:p1}
  \\
\min_{e_t, h_t}\mathop {\lim \sup }\limits_{M \to \infty } \frac{1}{M}\mathbb{E}\Big\{\sum_{t=1}^M X_{3,t}\Big\} \label{eq:p2}
  \\
  \text{s.t.}~
  E^{\min} \leq E_t \leq E^{\max}, \label{eq:ess_level}
  \\
  -d^{\max} \leq e_t \leq  c^{\max}, \label{eq:ess_min_max}
  \\
  E_{t+1} =
  \begin{cases}
    E_t +  \eta_c e_t \Delta t, & \text{if}~e_t > 0\\
    E_t + \frac{e_t}{\eta_d} \Delta t, & \text{else},
  \end{cases}
  \label{eq:ess_model}
  \\
  0\leq h_t\leq h^{\max} \label{eq:hvac_min_max},
  \\
  T^{\text{in}}_{t+1} = \mathcal{F}_T(T^{\text{in}}_{t}, T^{\text{out}}_t, h_t, \epsilon_t) \label{eq:building_model},
  \\
  g_t+p_t=h_t+e_t+l_t, \label{eq:power_balance}
\end{gather}
where the expectation operator $\mathbb{E}$ is imposed on uncertain system parameters, e.g., electricity prices $u^+/u^-$, PV generation $p_t$, outdoor temperature $T^{\text{out}}_t$, and non-shiftable load $l_t$. $X_{1, t}=u^+ g_t \Delta t|_{g_t > 0}+u^- g_t\Delta t|_{g_t \leq 0}$, $X_{2, t}=k_e |e_t|$, $X_{3,t}=[T^{\text{in}}_t - T^{\text{upp},\text{in}}]^++[T^{\text{low},\text{in}} - T^{\text{in}}_t]^+$ denote the energy transaction cost, ESS aging cost, and temperature deviation, respectively. Here, $u^+_t$, $u^-_t$, and $g_t$ denote the buying price, the selling price, and the transaction power at slot $t$, respectively. $\Delta t$ denotes the length of time slot $t$, \(k_e\) is an ESS depreciation coefficient, and $e_t$ denotes the ESS power at slot $t$. Specifically, $e_t>0$ denotes the charging power and $e_t<0$ denotes the discharging power. The operator $[\cdot]^+=\max\{\cdot,0\}$. $T^{\text{in}}_t$ is the indoor temperature at slot $t$; $T^{\text{low},\text{in}}$ and $T^{\text{upp},\text{in}}$ are the lower and upper bounds of comfort temperature, respectively.
In \eqref{eq:ess_level}, the range of energy level in the ESS is provided, where the minimum and maximum ESS energy level is $E^{\min}$ and $E^{\max}$, respectively. $E_{t}$ denotes the ESS energy level at slot $t$.
In \eqref{eq:ess_min_max}, the allowable charging/discharging power is provided, where $c^{\max}$ and $d^{\max}$ are the maximum charging power and discharging power, respectively.
\eqref{eq:ess_model} describes the ESS dynamics, where \(\eta_c\) signifies the charging coefficient and \(\eta_d\) denotes discharging coefficient.
\eqref{eq:hvac_min_max} describes the allowable range of HVAC input power $h_t$ and its maximum value is $h^{\max}$.
\eqref{eq:building_model} describes an implicit thermal dynamics model $\mathcal{F}_T$ for the smart home, which is the function of indoor temperature $T^{\text{in}}_t$, outdoor temperature $T^{\text{out}}_t$, HVAC input power $h_t$, and random thermal disturbance $\epsilon_t$. Without loss of generality, we assume that this implicit thermal dynamics model can be approximated by neural networks, similar to \cite{Natale2022}.
\eqref{eq:power_balance} denotes the power balance, where $p_t$ is the PV generation, and $l_t$ is the total non-shiftable loads.

\begin{figure*}
  \centering
  \includegraphics[width=0.95\textwidth]{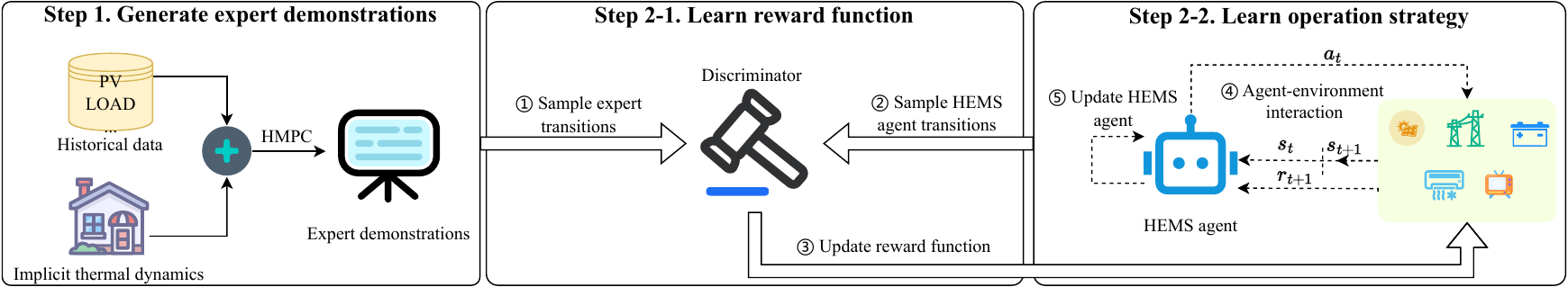}
  \caption{Architecture of the proposed SHEM method}
  \label{fig:algo}
\end{figure*}

\section{The Proposed Energy Management Method}\label{s3}
In this letter, we propose a SHEM method based on historical optimization decision-driven AIRL. In order to obtain historical optimization decisions in the presence of an implicit thermal dynamics model, a neural network-assisted HMPC is adopted. Then, such decisions are taken as expert demonstrations in the AIRL module, which aims to train a discriminator to distinguish expert demonstrations from transitions of the HEMS agent policy, and to update the HEMS agent's policy that confuses the discriminator. After finishing the above training process, a robust policy is obtained for online SHEM. The proposed HMPC-AIRL method has strong universality compared with existing model-based approaches and DRL-based approaches, since it does not require an explicit thermal dynamics model, prior or predicted information of uncertain parameters, or manually designed reward function.

The architecture of the proposed method is illustrated by Fig.~\ref{fig:algo}, where three blocks can be identified. In the first block, given a neural work-based implicit thermal dynamics model and historical data related to PV, price, outdoor temperature, and non-shiftable load, HMPC can be used to obtain near-optimal historical optimization decisions. Specifically, the upper-level MPC is utilized to optimize the HVAC input power \(h_t\) using a gradient descent method, and the optimization objective function is $\sum_{t=1}^Y X_{3,t}$, which can help to save energy by maintaining indoor temperatures close to the upper bound of the comfortable temperature range. Here, $Y$ denotes the slot number used for generating expert demonstrations. Then, taking the optimal HVAC input power vector as known values, the lower-level MPC is utilized to derive ESS decisions $e_t$ using the branch-and-bound method. More details of utilizing neural network-assisted HMPC to obtain historical optimization decisions of smart homes can be found in our previous work~\cite{yu2025coordinated}. Correspondingly, the obtained expert demonstrations, consisting of \(L\) samples, are expressed as
\begin{align}
  \tau^e = \left\{ (s_{1}, a_{ 1}, s_{1}'), (s_{ 2}, a_{2}, s_{2}'), \cdots,  (s_{L}, a_{L}, s_{L}')\right\},
\end{align}
where \(s_i, a_i, s_i'\) denote the \(i\)-th state, action and next state by taking action \(a_i\) at state \(s_i\), respectively.
Here, the state and action of the smart home environment \(\mathcal{E}\) at slot \(t\) is designed as \(s_t=(p_t, l_t, E_t, T^{out}_t, T_t^{in}, u_t, t')\) and $a_t=(e_t, h_t)$, respectively. Moreover, $t'=t \% 24$.

In the second block, a portion of expert demonstrations $\tau_n^e$ from $\tau^e$ is sampled. Moreover, the HEMS agent transitions $\tau_n^g$ are collected based on the interactions with the smart home environment $\mathcal{E}$ using the current policy $\pi_n^\theta$. Then, the two batches of transitions are labeled as $d_n$ (1 for expert transitions, 0 for HEMS agent transitions) and combined into $\tau_n^m$. Next, the combined transitions are fed into the discriminator to obtain the discriminator scores, and the score calculation formula is given by
\begin{align}
  \hat d_{n,i}=\frac{\exp\{r^\phi_{n}(s_i,a_i,s_i')\}}{\exp\{r^\phi_{n}(s_i,a_i,s_i')\}+\pi^\theta_{n}(a_i|s_i)},
  \label{eq:disc_score}
\end{align}
where $\hat d_{n,i}$ is the discriminator score of the $i$-th transition from the mixed transitions $\tau^m_n$, representing the confidence probability that this transition originates from expert demonstrations. A score closer to 1 indicates a higher probability that this transition originates from an expert.
$r^\phi_{n}$ is the reward function parameterized by $\phi$. $\pi^\theta_n$ is the policy function of the HEMS agent parameterized by $\theta$.
$(s_i,a_i,s_i')$ denotes the $i$-th transition. \(\pi^\theta_n(a_i|s_i)\) is the probability of taking \(a_i\) given state \(s_i\).
The structure of \(r^\phi_n\) exhibits a sophisticated design, comprising two sub-neural network components,
i.e., the reward approximator \(g_n^{\phi_1}\) and the shaping term \(h_n^{\phi_2}\)~\cite{fu2017learning}.
The calculation of \(r_n^\phi\) is given by
\begin{align}
  r_n^\phi(s_i,a_i,s_i') = g_n^{\phi_1}(s_i,a_i) + \gamma h_n^{\phi_2}(s_i') - h_n^{\phi_2}(s_i),
  \label{eq:r_cal}
\end{align}
where the additional shaping term \(h^{\phi_2}_n\) serves to minimize the influence of unwanted shaping on the reward approximator~\cite{fu2017learning}.
\(0 \ll \gamma < 1 \)  denotes a discount factor. Next, the discriminator scores and their corresponding labels are used to compute the discriminator loss, namely the binary cross-entropy loss between \(d_n\) and \(\hat d_n\):
\begin{align}\label{eq:disc_loss}
\mathcal{L}^D_{n} = \frac{1}{I}\sum\nolimits_{i=1}^{I}\left[ \mathcal{L}_i \right],
\end{align}
where $\mathcal{L}_i=-d_{n,i}\log \hat{d}_{n,i} - (1-d_{n,i})\log(1-\hat{d}_{n,i})$, \(I\) denotes the batch size. Subsequently, the loss is backpropagated to update the reward function from \(r_n^\phi\) to \(r_{n+1}^\phi\).

Once the current phase of the reward function updating is completed, the \(n\)-th training stage of the HEMS agent policy begins, as shown in the third block.
Specifically, a set of trajectories \(\mathcal{T}_n\) is collected from interaction episodes with the smart home environment \(\mathcal{E}\), using the current policy \(\pi^\theta_n\) and the updated reward function \(r_{n+1}^\phi\).
At each time step of every episode, the HEMS agent takes a control action and then receives the next state and reward signal calculated by \eqref{eq:r_cal} from \(\mathcal{E}\).
With the collected trajectories $\mathcal{T}_n$, the proximal policy optimization (PPO) algorithm is adopted to update the HEMS agent policy from \(\pi^\theta_n\) to \(\pi^\theta_{n+1}\) owing to its outstanding performance, stability, and ease of use. After a round of reward function updating and policy training, a new iteration is initiated. Moreover, these iterations continue until the termination condition is satisfied. For better illustration of the above-mentioned process, the details of the proposed method are provided in Algorithm~\ref{alg:algo}.

\begin{algorithm}
  \caption{The proposed HMPC-AIRL energy management method}
  \label{alg:algo}
  \SetKwInOut{KwIn}{Input}
  \SetKwInOut{KwOut}{Output}

  \let\oldnl\nl
  \newcommand{\nonl}{\renewcommand{\nl}{\let\nl\oldnl}}

  \KwIn{Given the implicit PCNN model \(\mathcal{M}\), smart home environment \(\mathcal{E}\), HEMS agent with initial policy \(\pi^\theta_0\), initial reward function \(r^\phi_{0}\), number of iterations \(N\)}
  \KwOut {\(\pi^\theta_{N}\)}

  Generate expert demonstrations \(\tau^e\) based on PHMPC energy management algorithm with \(\mathcal{M}\) in our previous work~\cite{yu2025coordinated} and historical data

  \For{$n \gets 0$ \KwTo $N-1$}{
    \nonl \# \emph{Learn the reward function of smart home} \\
    Sample a batch of expert transitions $\tau_{n}^e$ from \(\tau^e\)\\
    Collect a batch of HEMS agent transitions $\tau_{n}^g$ from \(\mathcal{E}\)  by the current policy \(\pi^\theta_n\)\\
    Label the two batches of transitions with \(d_n\) and mix them as \(\tau_n^m\)\\
    Compute the discriminator scores \(\hat d_n\) of \(\tau_n^m\) by \eqref{eq:disc_score}-\eqref{eq:r_cal}\\
    Compute the discriminator loss \(\mathcal{L}^D_{n}\) between \(d_n\) and \(\hat{d_n}\) by \eqref{eq:disc_loss}\\
    Backward \(\mathcal{L}^D_{n}\) to update the reward function as $r^\phi_{n+1}$\\

    \nonl \# \emph{Train the energy management policy} \\
    Collect some trajectories \(\mathcal{T}_n\) from \(\mathcal{E}\) with \(\pi^\theta_n\) and \(r_{n+1}^{\phi}\)\\
    Update the policy of the HEMS agent to \(\pi^\theta_{n+1}\) by PPO algorithm with \(\mathcal{T}_n\)
  }
\end{algorithm}

\section{Performance Evaluation}\label{s4}
Simulation experiments are conducted based on the real-world traces related to electricity price, PV generation, outdoor temperature, and non-shiftable load during the period from June 1, 2018 to August 31, 2018 from Pecan Street database\footnote{https://www.pecanstreet.org}. To simulate the actual indoor thermal dynamics, the explicit model in \cite{yu2023} is adopted so that the ideal upper-bound performance of the proposed method can be obtained. To obtain expert demonstrations, a neural network is used to approximate the above explicit model, and the purpose is to highlight that the proposed method is applicable to implicit thermal dynamics models represented by neural networks. Then, HMPC is used to obtain historical optimization solutions given the data in June and July. To indicate the data efficiency of the proposed method, two versions are adopted, i.e., 15-day demonstrations based on the data in June are adopted for \text{HMPC-AIRL-1 (HA1)} and 60-day demonstrations based on the data in June and July are adopted for \text{HMPC-AIRL-2 (HA2)}. For DRL-based benchmarks, the data in June and July are used for training and the data in August are used for testing. In addition, five benchmarks are adopted for comparisons, i.e.,
\begin{itemize}
  \item \text{DNLP}: it uses discontinuous non-linear programming (DNLP) solver to solve the optimization problem in Section~\ref{s2} with perfect parameter prediction information.
  \item \text{Rule}: it adopts the rule defined in~\cite{yu2023} for decisions.
  \item \text{DDPG}: it adopts explicit reward function (i.e., \mbox{\(r_t = -(X_{1,t} + X_{2,t}) - \beta_T X_{3,t}\)}, where \(\beta_T \) is a cost coefficient in $\$/^\circ \text{C}$) and uses deep deterministic policy gradient algorithm (DDPG) for HEMS agent training~\cite{YuJIOT2021}.
  \item \text{PPO}: it adopts the same reward function as \text{DDPG} and trains HEMS agent using PPO algorithm~\cite{Pinthurat2024}.
  \item \text{MaxEnt IRL}: it adopts MaxEnt~IRL~\cite{dey2023inverse} for agent training and uses the same demonstrations as \text{HA2} due to the difficulty in convergence when rule-based demonstrations are adopted.
\end{itemize}

\begin{figure}
  \centering
  \begin{subfigure}[T]{0.48\textwidth}
    \includegraphics[width=\linewidth]{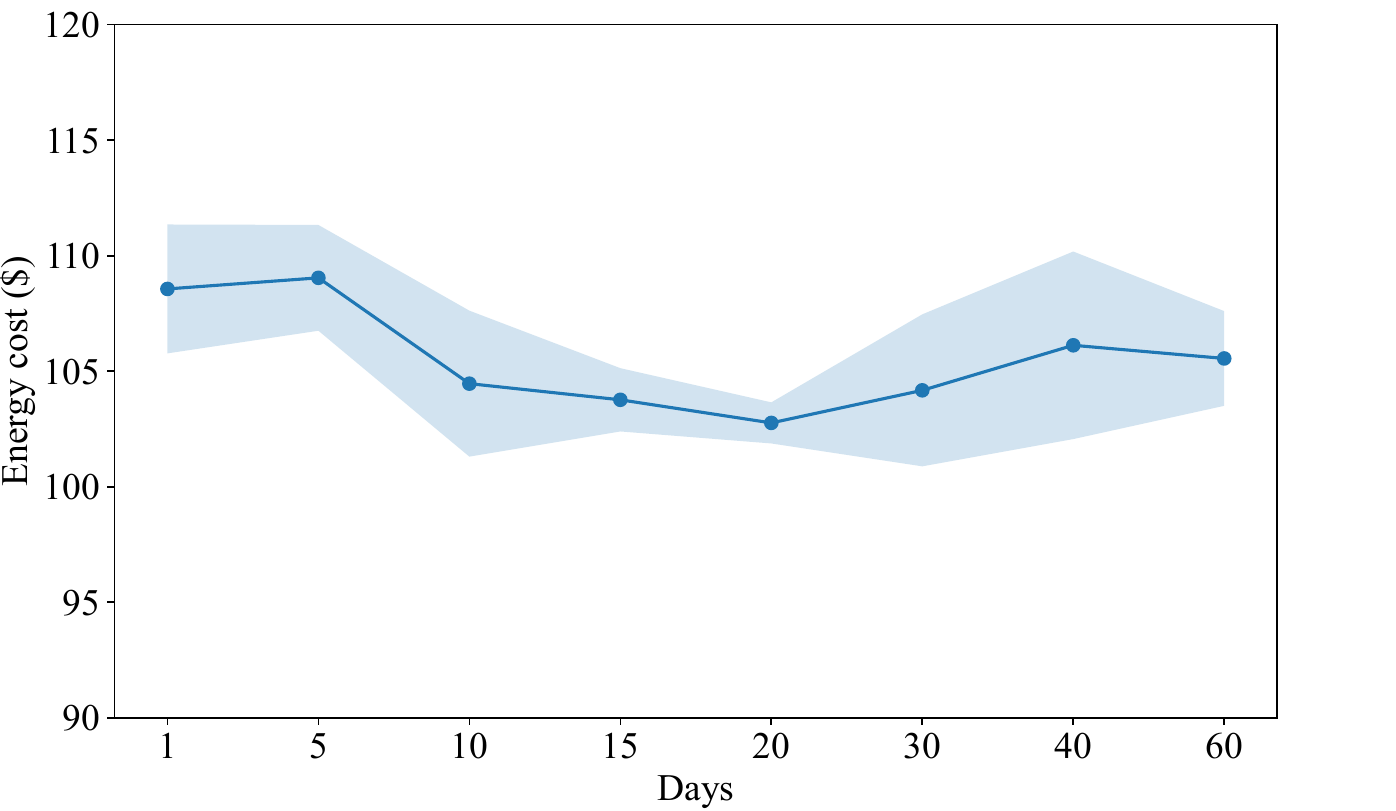} 
    \caption{Total energy cost}
  \end{subfigure}
  \begin{subfigure}[T]{0.48\textwidth}
    \includegraphics[width=\linewidth]{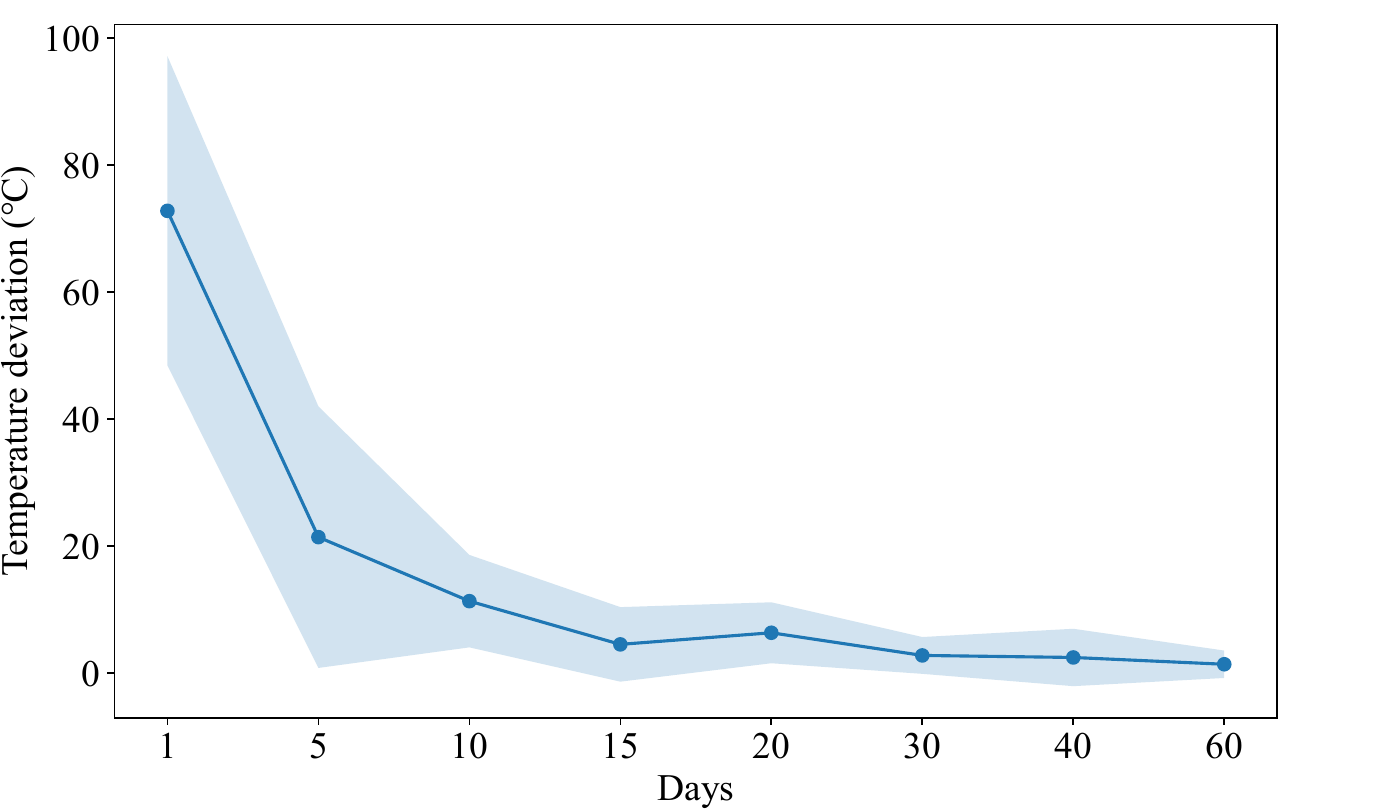} 
    \caption{Total temperature deviation}
  \end{subfigure}\\
  \begin{subfigure}[T]{0.48\textwidth}
    \includegraphics[width=\linewidth]{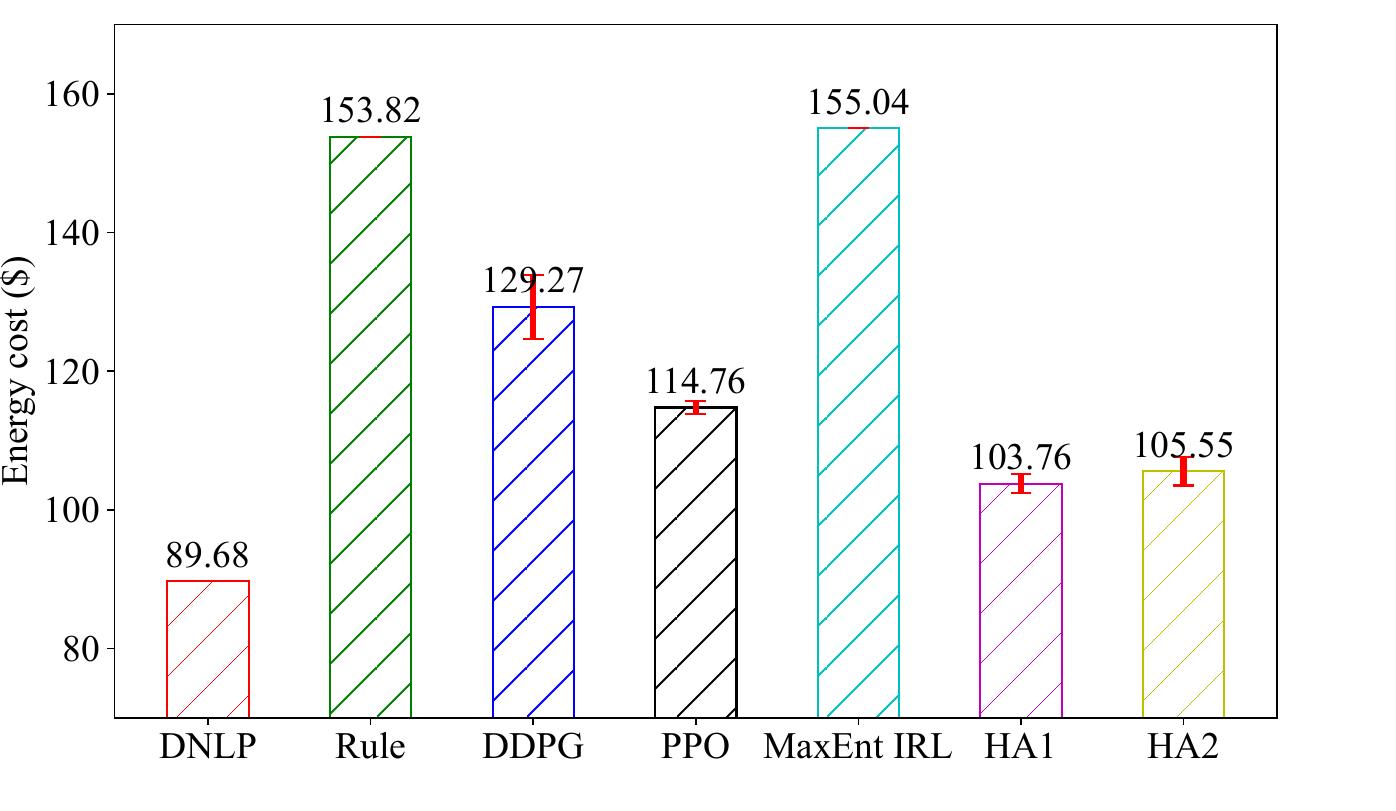} 
    \caption{Energy cost}
  \end{subfigure}
  \begin{subfigure}[T]{0.48\textwidth}
    \includegraphics[width=\linewidth]{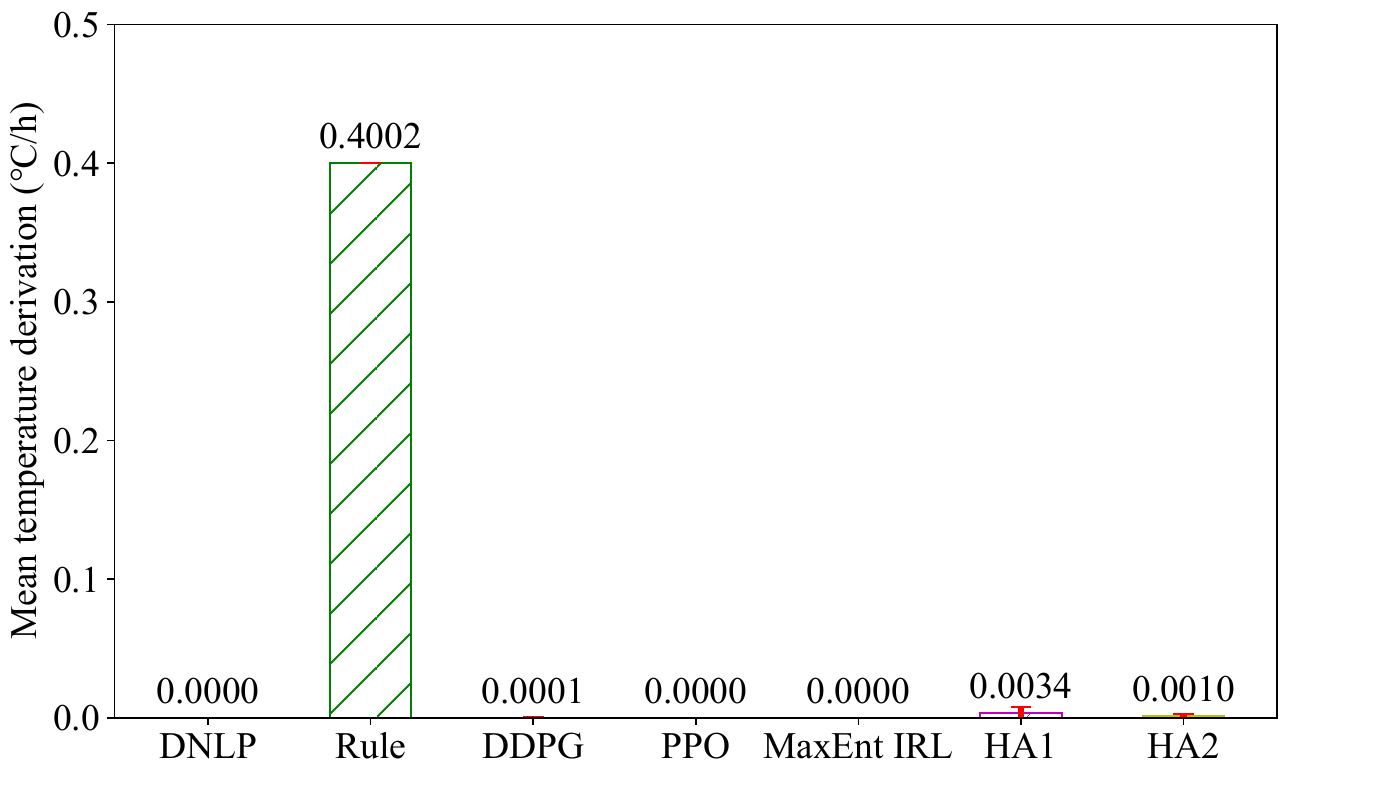} 
    \caption{Mean temperature derivation}
  \end{subfigure} \\
  \begin{subfigure}[T]{0.48\textwidth}
    \includegraphics[width=\linewidth]{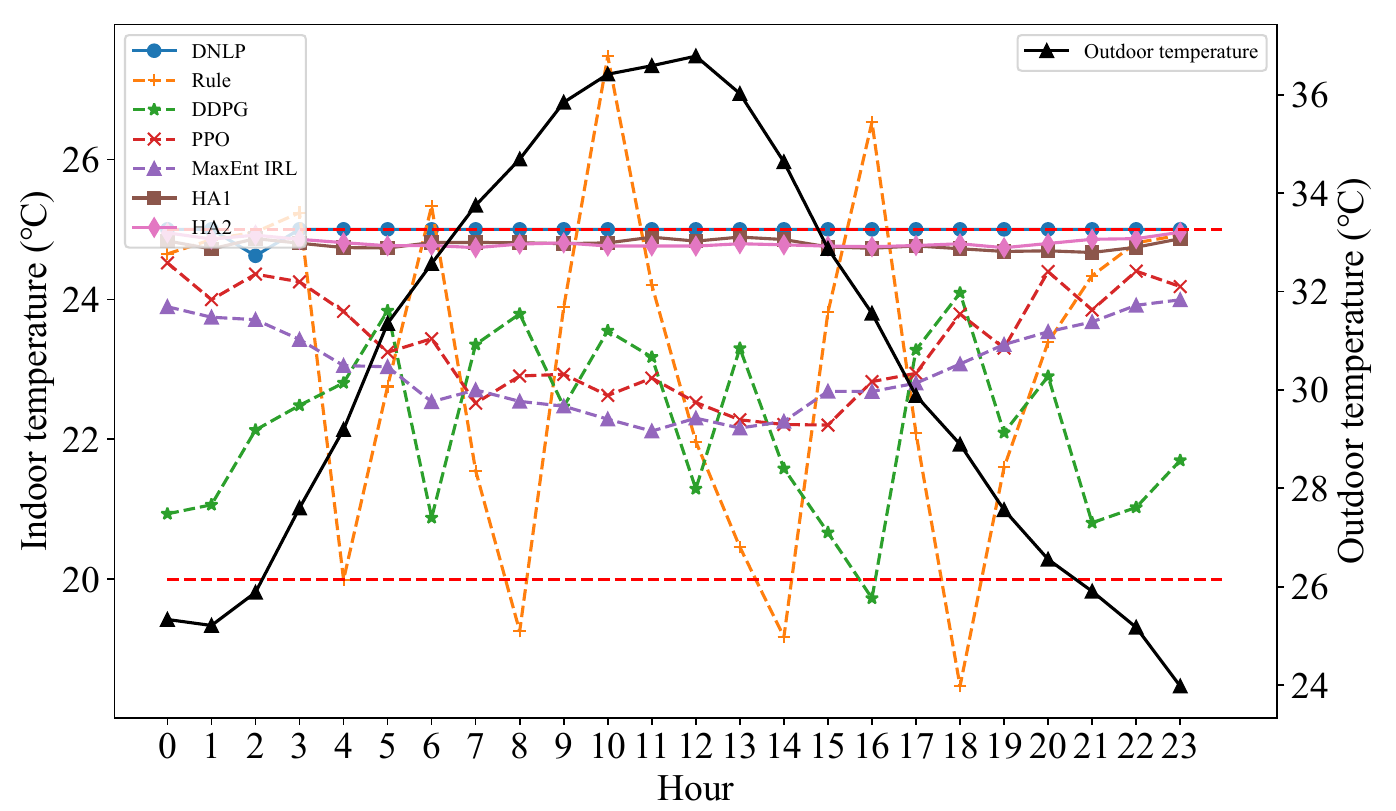} 
    \caption{Indoor temperature}
  \end{subfigure}
  \begin{subfigure}[T]{0.48\textwidth}
    \includegraphics[width=\linewidth]{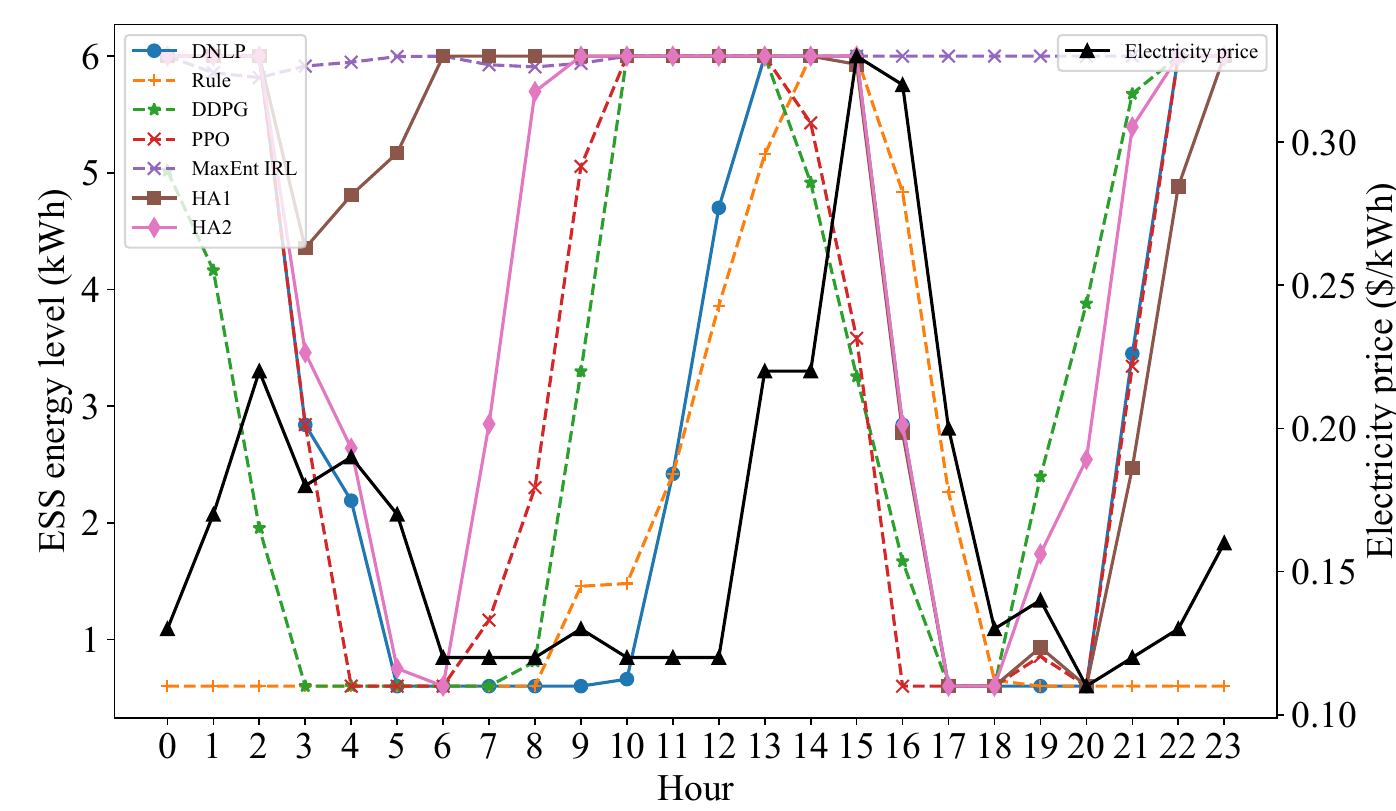} 
    \caption{ESS energy level}
  \end{subfigure}
  \caption{Performance comparisons}
  \label{fig:result}
\end{figure}


As shown in Figs.~\ref{fig:result}(a) and (b), with the increase in expert demonstration length, the total temperature deviation decreases rapidly. Moreover, the performance of the proposed method with 15-day expert demonstration is comparable to that of 60-day expert demonstration. As shown in Figs.~\ref{fig:result}(c) and (d), the proposed methods HA-1 and HA-2 achieve lower total energy cost (TEC) than other DRL-based benchmarks with negligible sacrifice in mean temperature deviation (MTD). For example, HA-1 can reduce TEC by 9.59\%-33.08\% compared with DRL-based benchmarks while using over 75\% less data. Compared with the Rule scheme, HA-1 reduces TEC and MTD by 32.51\% and 99.15\%, respectively. Although DNLP is impractical due to the requirement of accurate prediction information about uncertain parameters throughout the whole month, it offers a reference for the proposed method. According to calculations, the relative gap in TEC between DNLP and HA-1 is 15.70\%, which is far smaller than the 40\% in existing works~\cite{YuJIOT2021}. To explain the reason for the performance advantage of the proposed method, more results are provided in Figs.~\ref{fig:result}(e) and (f), where the proposed method maintains the indoor temperature close to the upper bound and controls the ESS to sell electricity during high-price periods and to purchase during low-price periods. Consequently, significant energy cost reduction is achieved. In summary, the proposed HMPC-AIRL method has advantages over existing DRL-based benchmarks in aspects of effectiveness and data efficiency.

\section{Conclusion}\label{s5}
This letter proposed a novel SHEM method based on HMPC and AIRL, which has no requirement of explicit thermal dynamic models, uncertainty parameter information, or manual reward function design. Simulation results demonstrated that it can reduce TEC by 9.59\% to 33.08\% while using over 75\% less data compared to other DRL-based benchmarks.

\end{document}